\documentclass[obeyspaces,
  journal=pasa,
  manuscript=research-paper, %
  year=202X,
  volume=YY,
]{cup-journal}

\usepackage{amsmath}
\usepackage{amssymb,microtype,siunitx,booktabs}
\usepackage[skip=0.5ex]{subcaption}
\usepackage{url}
\usepackage[nolist]{acronym}
\usepackage{hyperref} %

\begin{acronym}[AWGN]
\acro{AGB}{asymptotic giant branch}
\acro{ARC}{Australian Research Council}
\acro{ASKAP}{Australian Square Kilometre Array Pathfinder}
\acro{ATCA}{Australia Telescope Compact Array}
\acro{ATNF}{Australia Telescope National Facility}
\acro{CASDA}{CSIRO \ac{ASKAP} Science Data Archive}
\acro{CSIRO}{Australian Commonwealth Scientific and Industrial Research Organisation}
\acro{EM}{emission measure}
\acro{EMU}{Evolutionary Map of the Universe}
\acro{HLA}{Hubble Legacy Archive}
\acro{HST}{\textit{Hubble Space Telescope}}
\acro{LMC}{Large Magellanic Cloud}
\acro{MC}{Magellanic Clouds}
\acro{MIRIAD}[\textsc{Miriad}]{Multichannel Image Reconstruction, Image Analysis and Display}
\acro{MOST}{Molonglo Observational Synthesis Telescope}
\acro{MW}{Milky Way}
\acro{NRAO}{National Radio Astronomy Observatory}
\acro{ORC}{Odd Radio Circle}
\acro{PMN}{Parkes-MIT-NRAO}
\acro{PN}{Planetary Nebula}
\acro{rms}{root mean squared}
\acro{SED}{Spectral Energy Distribution}
\acro{SF}{star formation}
\acro{SN}{supernova}
\acro{SNR}{supernova remnant}
\acro{SUMSS}{Sydney University Molonglo Sky Survey}
\acro{WFC3}{Wide Field Camera 3}
\acro{WISE}{Wide-Field Infrared Survey Explorer}
\end{acronym}

\newcommand{\farcm}{\mbox{\ensuremath{.\mkern-4mu^\prime}}}%

\title{ASKAP--EMU radio continuum detection of planetary nebula NGC\,5189: the “Infinity” nebula}

\author{A. D. Asher}%
\affiliation{Western Sydney University, Locked Bag 1797, Penrith South DC, NSW 2751, Australia}
\alsoaffiliation{ATNF, CSIRO, Space and Astronomy, PO Box 76, Epping, NSW 1710, Australia}
\email[A. D. Asher]{albany.asher@csiro.au}

\author{Z. J. Smeaton}
\affiliation{Western Sydney University, Locked Bag 1797, Penrith South DC, NSW 2751, Australia}

\author{M. D. Filipovi\'c}
\affiliation{Western Sydney University, Locked Bag 1797, Penrith South DC, NSW 2751, Australia}

\author{A. M. Hopkins}%
\affiliation{School of Mathematical and Physical Sciences, 12 Wally's Walk, Macquarie University, NSW 2109, Australia}

\author{J. Th. van Loon}
\affiliation{Lennard-Jones Laboratories, Keele University, ST5 5BG, UK}

\author{T. J. Galvin}
\affiliation{ATNF, CSIRO, Space and Astronomy, PO Box 1130, Bentley, WA 6151, Australia}

\author{L. A. Barnes}
\affiliation{Western Sydney University, Locked Bag 1797, Penrith South DC, NSW 2751, Australia}

\received {dd Mmm YYYY}
\revised  {dd Mmm YYYY}
\accepted {dd Mmm YYYY}
\published{22 September 2020}


\begin{document}

\begin{abstract}
We report the radio continuum detection of well known Galactic \ac{PN} NGC\,5189, observed at 943\,MHz during the \ac{ASKAP} \ac{EMU} survey. Two detections of NGC\,5189 have been made during the survey, of better resolution than previous radio surveys. Both measurements of the integrated flux density are consistent with each other, at $S_{\rm 943\,MHz} = 0.33\pm0.03$\,Jy, and the spectral luminosity is $L_{\rm{943\,MHz}}$ = 8.89 $\times$ 10$^{13}$ W\,m$^{-2}$\,Hz$^{-1}$. Using available flux density measurements for radio-detections of NGC\,5189, we calculate a radio surface brightness at 1\,GHz and measure $\Sigma_{\rm 1~GHz}$ = 6.0 $\times$ 10$^{-21}$ W\,m$^{-2}$\,Hz$^{-1}$ sr$^{-1}$, which is in the expected range for Galactic PNe. We measure an apparent size of 3\farcm4\,$\times$\,2\farcm2 corresponding to physical diameters of 1.48\,pc\,$\times$\,0.96\,pc, and combine available radio observations of NGC\,5189 to estimate a spectral index of $\alpha$ = 0.12 $\pm$ 0.05. Hence, we agree with previous findings that NGC\,5189 is a thermal (free--free) emitting nebula. Additional measurements of the optical depth ($\tau = 0.00246$) and electron density ($N_{e} = 138~cm^{-3}$) support our findings that NGC\,5189 is optically thin at 943\,MHz. Furthermore, the radio contours from the \ac{ASKAP}--\ac{EMU} image have been overlaid onto a \ac{HST} Wide Field Camera 3 image, demonstrating that the radio morphology closely traces the optical. Notably, the contour alignment for the innermost region highlights the two envelopes of gas previously reported to be low-ionisation structures, which is considered a defining feature of post common--envelope PNe that surround a central Wolf-Rayet star.
\end{abstract}

\section{INTRODUCTION}\label{s:intro}

Planetary nebulae (PNe) are shells of gas that are shed during the terminal phase for the majority of stars between ${\sim}1-8$~M$_\odot$ \citep{Pottash1983,Gathier1986}, when they transition from the \ac{AGB} to the white dwarf stage \citep{Kwok2005}. The transitioning star ionises the surrounding shells and these remain visible for thousands of years, providing crucial insight into the recent mass loss rate, mechanism and nucleosynthesis of the \ac{AGB} star, as well as the evolution of the parent galaxy (particularly the chemical and star-forming evolution) \citep{Kwok2005,Gesicki2018,Crawford2015}. 

NGC\,5189 is a well known Galactic \ac{PN} located in the southern constellation Musca at RA (J2000) 13:33:31.8 and DEC (J2000) --65:58:29.8, at a distance of $\sim$1500 parsecs \citep{Chornay2021}. It was discovered by Scottish astronomer James Dunlop in 1826 \citep{Dunlop1828,Cozens2010} during observations from Parramatta, Australia, using a 9-inch reflector telescope\footnote{We refer the interested reader to James's original representation of NGC\,5189 (catalogued as object $\#$252) in \cite{Dunlop1828}.}. However, it was not considered a \ac{PN} until more recently by \cite{Evans1950} (an optical study), which they describe as “clearly not an ordinary planetary” and at the time they did not observe any central star. It was later determined to be a quadrupolar PN with multiple sets of “symmetrical condensations” (ansae, or knots) by \cite{Sabin2012} (both an optical and infrared study), and first identified as a binary system by \citealt{Manick2015}, where the central Wolf--Rayet [WO1] star \citep{Crowther1998} is in a binary orbit with potentially a main-sequence star or a white dwarf \citep{Manick2015}. 

NGC~5189 is renowned for its complex morphology and has been described by a number of authors (e.g. \citealt{Sabin2012,Danehkar2018,Aller2020}) as having filamentary and knotty structures, which can be attributed to the central binary system. A subsequent study by \cite{Bear2017}, however, suggests the structure may result from a ternary (triple) progenitor system \citep{Danehkar2018}. According to \cite{Phillips1983}, the symmetrical condensations that NGC\,5189 distinctly exhibits are the result of precessional torque induced by the companion star as it orbits the central star. 

\subsection{Radio Observations of PNe}

Radio continuum observations of PNe are not only useful for measuring the integrated flux densities and the associated spectral index, but also for tracing the thermal emission from the ionised plasma shells \citep{Hajduk2018} and determining the PN geometry and structures without the impediment of interstellar and atmospheric extinction \citep{Zijlstra1994}. While radio continuum surveys generally report the detection of weak thermal free--free emission from a PN \citep{Asher2024,Filipovic2009,Gathier1983,Milne1982,Pennock2021}, a few studies have reported non-thermal detections in the early photoionisation phase, as \ac{AGB} stars transition to the white dwarf stage \citep{Bains2009,Cerrigone2017,Perez2013,Suarez2015}. A notable example is the \cite{Hajduk2024} study regarding the radio continuum emission from Sakurai's Object (V4334 Sgr), the central star of a PN (CSPN) in the Sagittarius constellation. The authors observed the source between 2004 and 2023, finding non-thermal and highly variable emission to dominate between 2004 and 2017. This indicates Sakurai's object is a proto-PNe (or pre-planetary nebulae [PPN], \citealt{Kwok2001}) in the early photoionisation phase, of which the authors attribute the non-thermal emission to the shock interactions during mass-ejection events, as stellar winds from the \ac{AGB} star interact with the \ac{PN} shell \citep{Hajduk2024}.

\cite{Slee1965} were the first to observe NGC 5189 in the radio spectrum, at both 1420\,MHz and 2700\,MHz with the Parkes 64-m telescope. Subsequent Parkes radio detections (Table \ref{tbl:1}) were made between 1972 and 1982 at frequencies ranging from 2700\,MHz to 14.7\,GHz \citep{Aller1972,Milne1975,Milne1979,Milne1982}. 

More recent detections were made in 1990 at 4850\,MHz during the \ac{PMN} radio continuum survey \citep{Griffith1993,Wright1994,Condon1993} (conducted by the Parkes telescope and using the \ac{NRAO}-developed seven-beam receiver, \citealt{Condon1989}) and in 2007 at 843\,MHz during the second epoch Molonglo Galactic Plane Survey (MGPS-2) \citep{Murphy2007} -- the Galactic counterpart to the \ac{SUMSS} \citep{Bock1999, Mauch2003} by the \ac{MOST} \citep{Mills1981,Robertson1991}. 

We liken the appearance of the central region of NGC\,5189 to the lemniscate symbol, and hence our use of the colloquial moniker, \emph{Infinity} for this object.
Given the dearth of radio continuum measurements of \emph{Infinity} (hereafter used interchangeably with NGC\,5189) in recent years our primary goal has been to observe it and contribute up-to-date measurements using the \ac{ASKAP} radio telescope array. Therefore, in this paper we present the most recent radio continuum measurements for \emph{Infinity}, observed at 943\,MHz during the \ac{ASKAP} Evolutionary Map of the Universe (EMU) survey. 

The paper is structured as follows. In \S\ref{s:Data} we describe the \ac{ASKAP}--\ac{EMU} radio data and respective processes used to determine the new radio continuum measurements, in addition to detailing the corresponding \ac{HST} data used for comparative purposes. In \S\ref{s:ResultsDiscussion} we present the flux density, spectral index and associated results, and we summarise our findings in \S\ref{s:Summary}.

\vspace{5mm}

\section{DATA}\label{s:Data}

\subsection{ASKAP and the Evolutionary Map of the Universe Survey}
\label{s:EMU}

The Evolutionary Map of the Universe survey \citep{Norris2011,Hopkins2025} is a wide-field radio continuum galaxy survey being conducted by the \ac{ASKAP} radio telescope
\citep{Johnston2008,Hotan2021}, observing at the central frequency of 943\,MHz and a bandwidth of 288\,MHz \citep{Hopkins2025}. The EMU survey (ASKAP project AS201) began in May 2023 and will cover the entire southern sky by the scheduled time of completion in 2028 \citep{Hopkins2025}. The \ac{ASKAP} array consists of 36\,12-m antennas, with baselines up to 6\,km, and has a large instantaneous field of view (ranging between 15\,deg$^{2}$ at 1700\,MHz to 31\,deg$^{2}$ at 800\,MHz) \citep{Hotan2021}. 

\emph{Infinity} has been observed twice during the \ac{EMU} survey, with both observations using all 36 antennas over a duration of 10 hours. The respective scheduling blocks are SB53310 (observed on 30 September 2023) and SB62225 (observed on 7 May 2024). Both tiles are accessible via the CSIRO ASKAP Science Data Archive (CASDA)\footnote{https://research.csiro.au/casda/}. All \ac{EMU} data is processed through the ASKAPsoft pipeline \citep{Guzman2019}. 

Both images of \emph{Infinity} (see Figure~\ref{fig:HST_RGB}, top) have a resolution of 15.0\,$\times$\,15.0 arcseconds$^2$, as well as a Stokes~I image \ac{rms} noise value of 39\,$\mu$Jy beam$^{-1}$. For comparison, the median Stokes\,I image \ac{rms} noise value of each scheduling block (across the full tile) is 40\,$\mu$Jy beam$^{-1}$ (SB53310) and 44\,$\mu$Jy beam$^{-1}$ (SB62225), of which the sky area is 41.42 and 41.57\,deg$^{2}$ respectively. 

The data was predominantly analysed using the {\tt CARTA} \citep{Comrie2021} software, to establish the integrated flux densities for \emph{Infinity}. The flux density measurements are presented in Table~\ref{tbl:1}, in addition to measurements for the Parkes, \ac{PMN}, \ac{SUMSS} and MGPS-2 radio observations. For the \ac{PMN} observation, we relied upon the {\tt MIRIAD} \citep{Sault1995,Sault2011} and {\tt KARMA} \citep{Gooch1996} software packages to measure the integrated flux density and uncertainty level, as a consequence of statistical issues experienced when scaling the image in {\tt{CARTA}} (discussed further in \S\ref{s:ResultsDiscussion}).  

Numerous flux density measurements have been achieved and investigated in a number of \ac{ASKAP}--related papers, which we refer the interested reader to: including the \ac{LMC} \ac{ORC} J0624$-$6948 \citep{Filipovic2022,2025A&A...693L..15S}; the Galactic \ac{SNR} G278.94+1.35 (named “Diprotodon") by \citet{2024PASA...41..112F}; the discovery of a new, young, Galactic \ac{SNR} G329.9$-$0.5 (“Perun") by \cite{Smeaton2024}; the detection of a large and low surface brightness Galactic \ac{SNR} G288.8$-$6.3 (“Ancora") by \citep{Filipovic2023} (with a follow-up study by \cite{Burger2024}); and others (e.g. \citealt{Lazarevic2024,Smeaton_2024,Filipovic2025}).

\subsection{HST Data}
\label{s:HST}

\begin{figure*}[ht!]
\centerline{\includegraphics[width=0.85\textwidth, keepaspectratio]{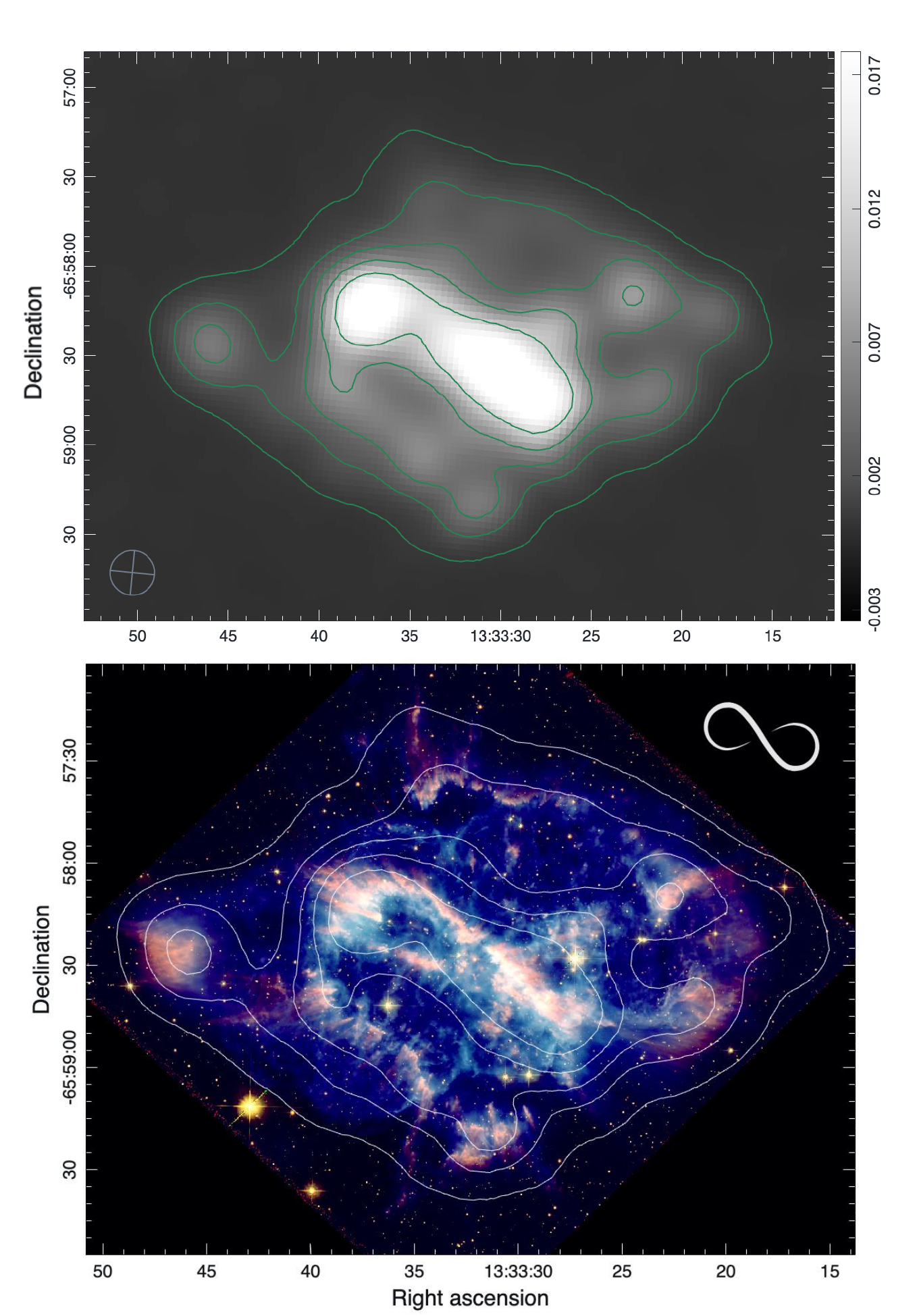}}
\caption{{\bf Top: } \ac{ASKAP}--\ac{EMU} radio image of \emph{Infinity} at 943\,MHz with an \ac{rms} noise level of 39\,$\mu$Jy beam$^{-1}$. The contours are at levels of 10, 50, 100, 175, and 200$\sigma$. The synthesised beam size is shown in the bottom left corner. {\bf Bottom: } RGB image of \emph{Infinity} from HST images, overlaid with \ac{ASKAP}--\ac{EMU} radio image contours. Red is the F673N filter (a narrow-band filter centred at 676.59\,nm), green is the F606W filter (a wide V band filter centred at 588.92\,nm), and blue is the F502N filter (a narrow-band filter centred at 500.96\,nm). All images are from the WFC3 instrument.}
\label{fig:HST_RGB}
\end{figure*}

We also use the available \ac{HST} data from the \ac{HLA}\footnote{https://hla.stsci.edu/} to compare the optical and radio properties. The images are from project 12812 (PI: Z. Levay), taken on 6 July 2012 using the \ac{WFC3} instrument\footnote{Refer to the WFC3 Instrument Handbook \citep{Marinelli2024} for further information on UVIS imaging with WFC3: https://hst-docs.stsci.edu/wfc3ihb} \citep{Leckrone1998}. We make use of three images to create a composite RGB image (Figure~\ref{fig:HST_RGB}, bottom), specifically the F673N (red), a narrow-band filter centred at 676.59\,nm with a width of 11.78\,nm (which covers the [S\,{\sc ii}] doublet), F606W (green), a wide V band filter centred at 588.92\,nm with a width of 218.92\,nm (which covers the [O\,{\sc iii}] doublet [both lines] and the H$\alpha$ line), and F502N (blue), a narrow-band filter centred at 500.96\,nm with a width of 6.53\,nm (which covers the red [O\,{\sc iii}] line). We use these images for comparison with radio, and an in-depth optical emission line analysis using the same data can be found in \cite{Danehkar2018}.

\section{RESULTS AND DISCUSSION}\label{s:ResultsDiscussion}

Flux density measurements of Galactic PNe are too varied to derive a suitable expectation for a single PN, which is not surprising given the variation in PN distances (the exception to this issue are the \ac{MC} PNe, which remain unresolved, \citealt{Pennock2021}). Based on the variation, we contribute supplementary measurements of the spectral luminosity\footnote{We assume an isotropic source when using the formula $L_{v} = 4\pi d^{2}S_{v}$, where $\nu$ is the \ac{ASKAP}--\ac{EMU} frequency and \emph{d} is the \cite{Chornay2021} distance to \emph{Infinity} in metres \citep{Marr2015}.} at 943\,MHz, %
and the distance-independent radio surface brightness, $\Sigma$, at 1\,GHz (a standard scaling frequency for determining surface brightness values, e.g. \citealt{Cotton2024}) and determine if the value falls in the expected range for Galactic PNe. 

We also explore a historical comparison of flux density values. Historically, the values for \emph{Infinity} have been measured between 0.215\,Jy at 4850\,MHz \citep{Griffith1993,Wright1994} and 0.459\,Jy at 14700\,MHz \citep{Milne1982}. Using {\tt CARTA} and the polygon tool to define the outline of \emph{Infinity}, we measured an integrated flux density of $S_{\rm 943\,MHz} =$ 0.333\,Jy (SB53310) and $S_{\rm 943\,MHz} =$ 0.334\,Jy (SB62225). The {\tt CARTA} software does not specifically provide uncertainty levels, however the {\tt MIRIAD IMFIT} task does, available from the {\tt MIRIAD} \citep{Sault1995,Sault2011} software package developed by the \ac{ATNF}. Two separate assessments of the \emph{Infinity} images indicate an uncertainty level of $\sim$10$\%$ for both images. The \ac{ASKAP} measurements are presented in Table \ref{tbl:1}, in addition to the historical flux density values and their respective uncertainties. 

In respect to the apparent size of \emph{Infinity}, using {\tt{CARTA}} and retaining the polygon outline, we measured the angular dimensions of 3\farcm4 $\times$ 2\farcm2. Using the \cite{Chornay2021} distance of $\sim$1500\,pc in our calculation, this corresponds to a physical size of 1.48\,pc $\times$\,0.96\,pc. Using the same distance of $\sim$1500\,pc, we measure the spectral luminosity, $L_{\rm 943\,MHz}$ = 8.89\,$\times\,10^{13}$ W\,m$^{-2}$\,Hz$^{-1}$.

We also employ the mean of the angular dimensions to determine the surface brightness of \emph{Infinity} at 1\,GHz, of which we measure 6.0$\times10^{-21}$ W\,m$^{-2}$\,Hz$^{-1}$\,sr$^{-1}$. This value falls within the expected range of radio surface brightnesses for Galactic PNe (between $\sim10^{-24} - 10^{-16}$ W\,m$^{-2}$\,Hz$^{-1}$\,sr$^{-1}$, \citealt{Leverenz2016}), again placing confidence in the measurements. 

Additionally, we measured the integrated flux densities of \emph{Infinity} in the \ac{SUMSS} and \ac{PMN} surveys. The reason for measuring \ac{SUMSS} is that the integrated flux density value for \emph{Infinity} is not present in the respective catalogue. For the \ac{PMN} observation, there are two different measurements presented in the catalogue (where \emph{Infinity} is listed in the final source catalogue, “Table 2” in  \cite{Wright1994} and Table 2 in the online point source catalogue \footnote{The \ac{PMN} point source catalogues and survey maps can be accessed at: https://www.parkes.atnf.csiro.au/observing/databases/pmn/pmn.html}, as per its J2000--derived source name, J1333--6558). The two measurements are $S_{\rm 4850\,MHz} =$ 0.309$\pm$0.017\,Jy and $S_{\rm 4850\,MHz} =$ 0.215\,Jy, respectively. Initially, we considered that the latter measurement may have resulted from measuring a steep spectral source, as the value is not in close agreement with the majority of historical observations. However, \cite{Wright1994} explain that a few complex objects were larger than the beam size and led to multiple measurements of those sources being included in Table 2. The measurement of 0.215\,Jy was determined using a “General-width fit” \citep{Griffith1993}, designed for detecting and measuring extended sources, where the maximum and minimum widths of a source are normalised to the beam size of 4$\farcm$2. This method may explain the discrepancy between the two measurements. Regardless, we choose to  measure the \ac{PMN} \emph{Infinity} detection again. 

We accessed the respective FITS files via SkyView\footnote{https://skyview.gsfc.nasa.gov} and analysed accordingly in {\tt{CARTA}}, assuming a 10$\%$ uncertainty level as described and investigated in the aforementioned \ac{ASKAP}--related studies (see \S\ref{s:EMU}). For the 843\,MHz image in the SUMSS survey, we measured an integrated flux density of 0.245$\pm$0.025\,Jy. We note that {\tt{CARTA}} did not generate the integrated flux density directly (due to missing information in the image file header, most notably the restoring beam and frequency), but rather we used {\tt{CARTA's}} \emph{Sum} flux value of the selected region, 4.644\,Jy, to determine the integrated flux density using the following formula \citep{1996PhDT........32F}:
\begin{eqnarray}
    S_\nu = \frac{\rm Sum}{1.133(\frac{BS}{PS})^{2}}\,,
\end{eqnarray}
where $S_{\nu}$ is the integrated flux density at a given frequency, \emph{Sum} is the flux value of the selected region (in Jy), \emph{BS} is the beam size (for \ac{SUMSS} this is 45$^{\prime\prime}\times45^{\prime\prime}$) and \emph{PS} is pixel size (for \ac{SUMSS} this is 11$^{\prime\prime}\times11^{\prime\prime}$).

For the 4850\,MHz image in the PMN survey, limited header information in the image file prevented measurements using {\tt{CARTA}}. Therefore, we employed {\tt MIRIAD IMFIT} and determined a flux density value of 0.310$\pm$0.004\,Jy, which is in close agreement with the value listed in the final source catalogue in \cite{Wright1994}. The \ac{PMN}\footnote{For the original \ac{PMN} \emph{Infinity} detection, we present the final source catalogue value \citep{Wright1994} in Table \ref{tbl:1} only.} and \ac{SUMSS} measurements are also presented in Table \ref{tbl:1}.

\begin{table*}%
\centering
\caption{We present the available flux density measurements for radio-detections of NGC\,5189 (Column\,4), in chronological order. %
$^{\ast}$Year of catalogue release, excepting \ac{ASKAP}--\ac{EMU}, which are years of observation. $^{\dag}$Average integrated flux density measurements, based on the average of 8 or 9 scans of \emph{Infinity}, conducted by the Parkes telescope (8 scans for observations at 2700\,MHz and 9 scans at 1420\,MHz). $^{\ddag}$New integrated flux density measurements of existing radio data.} 
\vspace{2mm}
\begin{tabular}{ccccccl}
\hline
&Year$^{\ast}$ &Telescope/Survey &$\nu$ & $S\pm\Delta S$ & Beam Size &References\\
&  &  & (MHz) & (Jy) & \\
\hline
&1965 &Parkes 64-m &1420 &0.410$\pm$0.021$^{\dag}$ & 14\farcm0 $\times$ 14\farcm0 &Slee \& Orchiston (1965)\\
&1965 &Parkes 64-m &2700 &0.360$\pm$0.018$^{\dag}$ & 7\farcm5 $\times$ 7\farcm5 &Slee \& Orchiston (1965)\\
&1972 &Parkes 64-m &2700 &0.330$\pm$0.100 & 8\farcm1 $\times$ 8\farcm1 &Aller \& Milne (1972)\\
&1975 &Parkes 64-m &5000 &0.366$\pm$0.025 & 4\farcm5 $\times$ 4\farcm5 
&Milne \& Aller (1975)\\
&1979 &Parkes 64-m &5000 &0.413$\pm$0.045 & 4\farcm5 $\times$ 4\farcm5 
&Milne (1979)\\
&1982 &Parkes 64-m &14700 &0.459$\pm$0.023 & 2\farcm1 $\times$ 2\farcm1 
&Milne \& Aller (1982)\\
&1994 &Parkes--MIT--NRAO &4850 &0.309$\pm$0.017 & 4\farcm2 $\times$ 4\farcm2 
&Wright et al. (1994)\\
&1994 &Parkes--MIT--NRAO &4850 &0.310$\pm$0.004$^{\ddag}$ & 4\farcm2 $\times$ 4\farcm2 
&This paper, based on Wright et al. (1994).\\
&2003 &MOST--SUMSS &843 &0.245$\pm$0.025$^{\ddag}$ & 45 $\times$ 45 csc $| \delta |$ arcsec$^{2}$
&This paper, based on Mauch et al. (2003).\\
&2007 &MOST--MGPS--2 &843 &0.307$\pm$0.012 & 45 $\times$ 45 csc $| \delta |$ arcsec$^{2}$
&Murphy et al. (2007)\\
&2023 &ASKAP--EMU &943 &0.333$\pm$0.033 & 15$^{\prime\prime} \times 15^{\prime\prime}$
&This paper.\\
&2024 &ASKAP--EMU &943 &0.334$\pm$0.033 & 15$^{\prime\prime} \times 15^{\prime\prime}$
&This paper.\\
\hline
\end{tabular}
\label{tbl:1}
\end{table*}

\begin{figure*}[ht!]
\centering
\vspace{3mm}
\begin{tabular}{c c }
        \textbf{Parkes--MIT--NRAO} & \textbf{ASKAP--EMU} \\
        \includegraphics[angle=0, trim=0 0 0 0, width=.46\textwidth]{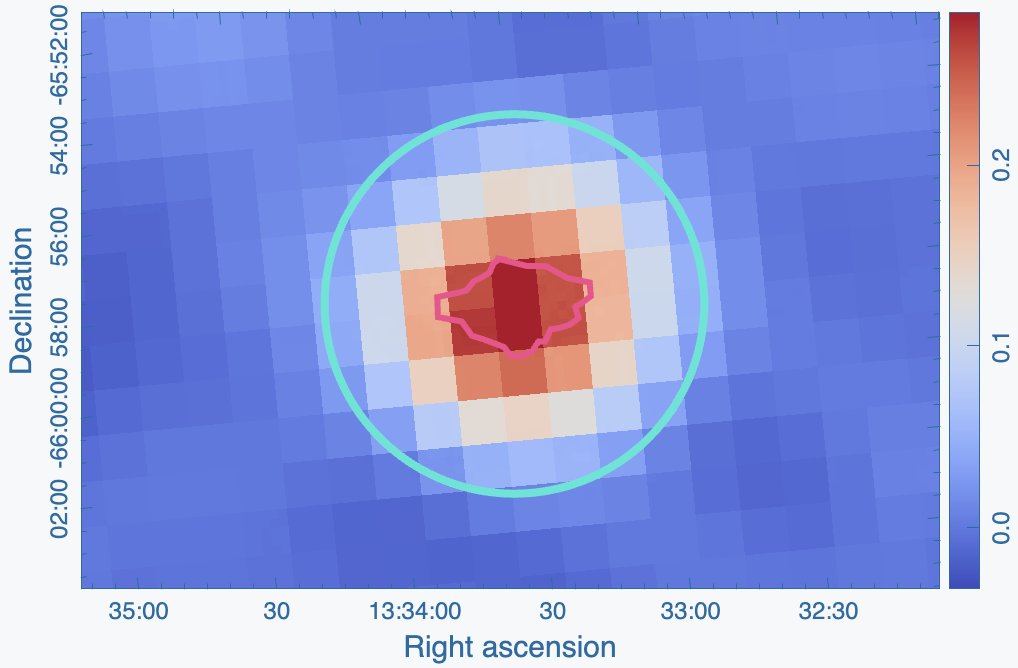} &
        \includegraphics[angle=0, trim=0 0 0 0, width=.46\textwidth]{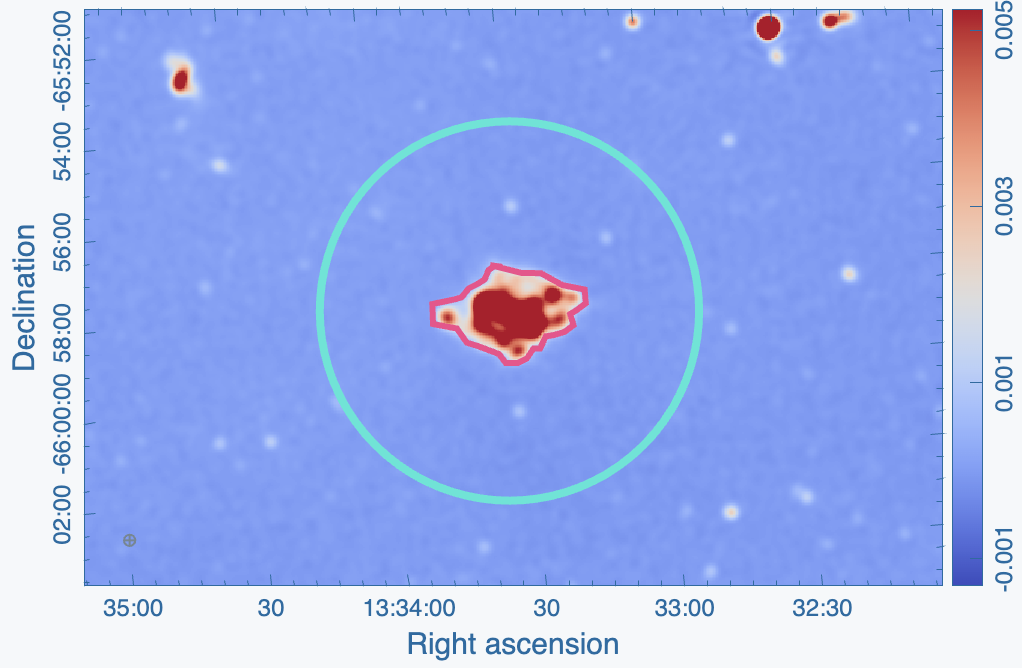}\\
    \end{tabular}
\caption{A comparison of \emph{Infinity} observations: the cyan ellipse in both the \ac{PMN} and \ac{ASKAP}--\ac{EMU} images represents the region that was measured during the \ac{PMN} survey \citep{Griffith1993,Wright1994,Condon1993}. The central pink polygon in both images represents the \emph{Infinity} \ac{PN} from the recent \ac{ASKAP} observations.}
\label{fig:PMN}
\end{figure*}

Importantly, when analysing the \ac{PMN} \emph{Infinity} observation in {\tt CARTA}, we created an ellipse region to represent the boundary of the second measurement (as per the 4\farcm2 beam size) and applied to both the \ac{ASKAP}--\ac{EMU} and PMN images. %
In Figure~\ref{fig:PMN}, the normalised PMN region of 4\farcm2\,$\times$\,4\farcm2 is indicated by the cyan ellipse in both images (to the left is \ac{PMN} and to the right is \ac{ASKAP}--\ac{EMU}, where both intensity scales are measured in Jy/beam). For comparison, the central region observed by \ac{ASKAP} is indicated by the pink polygon. In the \ac{ASKAP}--\ac{EMU} image, there are a number of background sources that fall within the PMN-catalogued region. Upon assessment of the four most distinct background sources (which have a combined integrated flux value of $\sim$0.0024\,Jy), it is reasonable to conclude that they have a negligible effect, despite being included in the measurement. Additionally, we would expect steep spectral sources observed at 4850\,MHz to be fainter than observations at $\sim$1\,GHz. The reason the second measurement is lower is therefore unclear.

\begin{figure*}[ht!] %
\centerline{\includegraphics[width=0.95\textwidth, keepaspectratio]{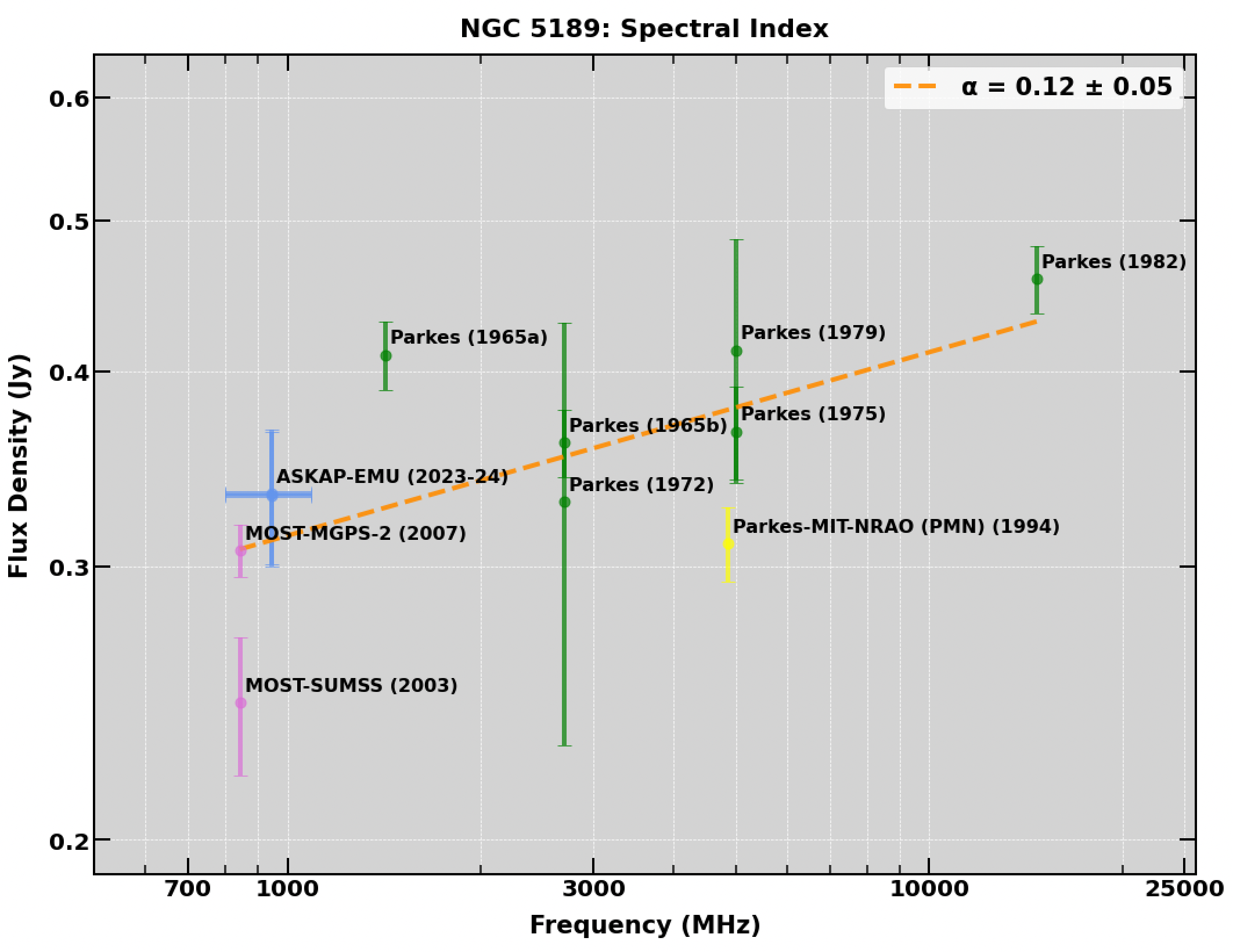}}
\caption{Using 11 of the 12 available radio data points (labelled according to the respective telescope and year of the associated paper or observation), we calculated the radio spectral index for \emph{Infinity} and determined $\alpha$ = 0.12$\pm$0.05, represented by the dashed orange line. Both axes are log scale. For the two ASKAP-EMU data points coloured blue, the survey bandwidth of 288\,MHz has been marked with a horizontal line of the same colour.} 
\label{fig:specindex}
\end{figure*}

In Figure~\ref{fig:specindex}, we estimate the spectral index using 11 of the 12 radio data points listed in Table \ref{tbl:1}. Only one of the \ac{PMN} measurements can be included in the calculation, to correspond with one observation, of which we employ the value listed in the final source catalogue \citep{Wright1994}. We estimate a spectral index\footnote{Spectral index is defined as $S\propto\nu^{\alpha}$, where $S$ is integrated flux density, $\nu$ is observing frequency, and $\alpha$ is spectral index.} of $\alpha$ = 0.12$\pm0.05$. Thermal free--free emission is typically associated with spectral indices of $\alpha = 0 - 2$ and non-thermal with $\alpha <$ --0.5 \citep{Ridpath2018}. Accordingly, we infer that NGC\,5189 is a thermal (free--free) emitting nebula. This finding agrees with previous observations (e.g. \citealt{Aller1972,Milne1982}).

Inspection of the data in Figure~\ref{fig:specindex} shows that the three data points at lower frequencies ($<$1000\,MHz in this instance, i.e. SUMSS, MGPS-2, ASKAP--EMU) indicate a small degree of brightening over a 21--year period (from 0.245\,Jy at 843\,MHz [SUMSS] up to 0.334\,Jy at 943\,MHz [ASKAP--EMU]), although this is marginal. Based on the combination of age, differing flux scales and measurement techniques, the Parkes (1965a) and PMN (1994) data points are likely to be outliers; with emphasis on the differing flux density scales, as both sets of observations were determined using the radio galaxy Hydra-A (PKS B0915--118) to calibrate the absolute flux density scale \citep{Slee1965,Condon1993}. For comparison, \ac{ASKAP} employs radio galaxy PKS B1934--638 to calibrate the absolute flux density scale \citep{Hotan2021}. 

The data point at 14.7\,GHz \citep{Milne1982} is slightly higher than the others, yet in  agreement with an earlier Parkes observation at 5\,GHz \citep{Milne1979}, and certainly within the uncertainty level of the 1979 observation. \cite{Milne1982} compared the 14.7\,GHz and earlier 5\,GHz flux density measurements \citep{Milne1979} for 236 confirmed PNe (including \emph{Infinity}) and found there to be notable scattering in the diagram. They mostly attribute this to the fixed error when measuring flux densities, in addition to the fractional error associated with calibration.

The remaining data points at higher frequencies ($\ge$2700\,MHz) show more consistency with little indication of variation with time. There is a marginal drop in flux density between the 1965b and 1972 Parkes observations at 2700\,MHz, from 0.360\,Jy to 0.330\,Jy, respectively, but a similarly sized marginal increase in flux density at 5000\,MHz between the 1975 and 1979 Parkes observations, although in both cases these are within the estimated uncertainties. The gap between the earliest (Parkes) and most recent (\ac{ASKAP}) observations is six decades, and any temporal variation in flux density or spectral index on such a timeframe for this class of object is likely to be small. 

According to \cite{Hajduk2018}, flux evolution is not only dependent on factors such as age (where, for example, electron densities will generally decline monotonically as the PNe expands, \citealt{Zhang2004}) but also the central star's evolution, which vary in their progress and can contribute to variation in flux measurements.

Given the spectral index value of \emph{Infinity} lies closer to the optically thin case of $\alpha = -0.1$ than the optically thick at $\alpha = 0.6-2.0$ (where more light is absorbed) \citep{Taylor1987,Gruenwald2007}, we also infer that \emph{Infinity} is predominantly within the optically thin regime for observations $\gtrsim$1000\,MHz. 

To explore this result further, we rewrite the Rayleigh-Jeans approximation\footnote{The approximation is suitable to use as it holds for frequencies $\frac{\nu}{\text{GHz}} \ll 21\left(\frac{T}{K}\right)$ \citep{Wilson2013}.} so we can measure the optical depth, $\tau$, which is proportional to the \ac{EM} (defined as the integral of the electron density, $N_{e}$, along the line of sight in an emission nebula, \citealt{Wilson2013}):
\begin{eqnarray}
   S_{\nu} = \frac{2kv^{2}}{c^{2}} \; T_{B}\Delta\Omega \,,
\label{eqnrayleigh}
\end{eqnarray}
\noindent where \emph{S}$_{\nu}$ is the integrated flux density at a specific frequency, \emph{k} is the Boltzmann constant, \emph{c} is the speed of light, \emph{T}$_{B}$ is the brightness temperature and $\Delta\Omega$ is the angular area (measured as an elliptical area in this instance, equating to 4.971\,$\times$\,10$^{-7}$\,sr). We rearrange Eq. \ref{eqnrayleigh} to solve for $T_{B}$:
\begin{eqnarray}
   T_{B} = \frac{S_{\nu}c^{2}}{2k\nu^{2}\Delta\Omega}  \,.
\label{eqntb}
\end{eqnarray}
\noindent We measure $T_{B} = $ 24.6\,K. Brightness temperature can also be expressed as $T_{B} = T(1 - e^{-\tau})$, which we rearrange to solve for $\tau$:
\begin{eqnarray}
   \tau = -ln\left(1 - \frac{T_{B}}{T}\right)\,.
\label{eqnoptdepth}
\end{eqnarray}
\noindent We measure $\tau = $ 0.00246. Since $\tau \ll 1$, this reinforces the conclusion that \emph{Infinity} is optically thin at 943\,MHz. To determine the \ac{EM} and $N_{e}$ values, we employ the \cite{Mezger1967} approximation for the free-free opacity $\tau$:
\begin{equation}
\resizebox{0.875\columnwidth}{!}{$\tau = 3.28 \times 10^{-7} \left(\frac{T}{10^{4}K}\right)^{-1.35} \left(\frac{\nu}{GHz}\right)^{-2.1} \left(\frac{EM}{\text{pc cm}^{-6}}\right)$} \,,
\label{eqntau}
\end{equation}
\noindent where for \emph{T} we use the canonical value (10$^{4}$~K) \citep{Bojicic2021}, and $\nu$ is the frequency of the \ac{ASKAP}--\ac{EMU} survey in GHz (0.943). We then rearrange Eq. \ref{eqntau} and solve for \ac{EM} specifically:
\begin{eqnarray}
   EM = \frac{\tau}{3.28 \times 10^{-7} \left(\frac{T}{10^{4}K}\right)^{-1.35}\left(\frac{\nu}{GHz}\right)^{-2.1}}\,.
\end{eqnarray}
\noindent We measure $EM =$ 6630\,pc cm$^{-6}$. From the following \ac{EM} formula \citep{Wilson2013} we can now rearrange and measure $N_{e}$:
\begin{eqnarray}
   EM = \int_{0}^{s}\left(\frac{N_{e}}{cm^{3}}\right)^{2} \; d\left(\frac{s}{\text{pc}}\right)\\\therefore \frac{N_{e}}{cm^{3}} = \sqrt \frac{EM}{\left(\frac{s}{pc}\right)}\,.
\label{EM}
\end{eqnarray}
\noindent where overall this represents the integral of the electron density squared (\emph{N}$_{e}^{2}$) at a depth (\emph{s}) measured in pc. To determine the depth, we assume a cylindrical geometry and a path length equal to the smallest diameter of 2\farcm2. We measure $N_{e} =$\,138\,cm$^{-3}$. Optically thin PNe have electron densities $<$5000\,cm$^{-3}$, whereas for optically thick PNe the electron densities are $>$6000\,cm$^{-3}$ \citep{Barlow1987}. Therefore, our result supports our earlier findings that \emph{Infinity} is optically thin at 943\,MHz. 

Notably, from Figure~\ref{fig:HST_RGB} it is evident that the radio continuum emission closely follows the optical. Specifically, the radio emission closely traces the F606W filter. This is possibly due to the H$\alpha$ emission line at 656.3\,nm, which has a throughput close to the peak of the integrated system, $\sim$29\%. H$\alpha$ is related to the distribution and density of thermally-emitting ionised gas \citep{Tacchella2022}. Therefore, it provides valuable insight into the structure and morphology of PNe. 

\begin{figure}
\centerline{\includegraphics[width=0.85\columnwidth, keepaspectratio]{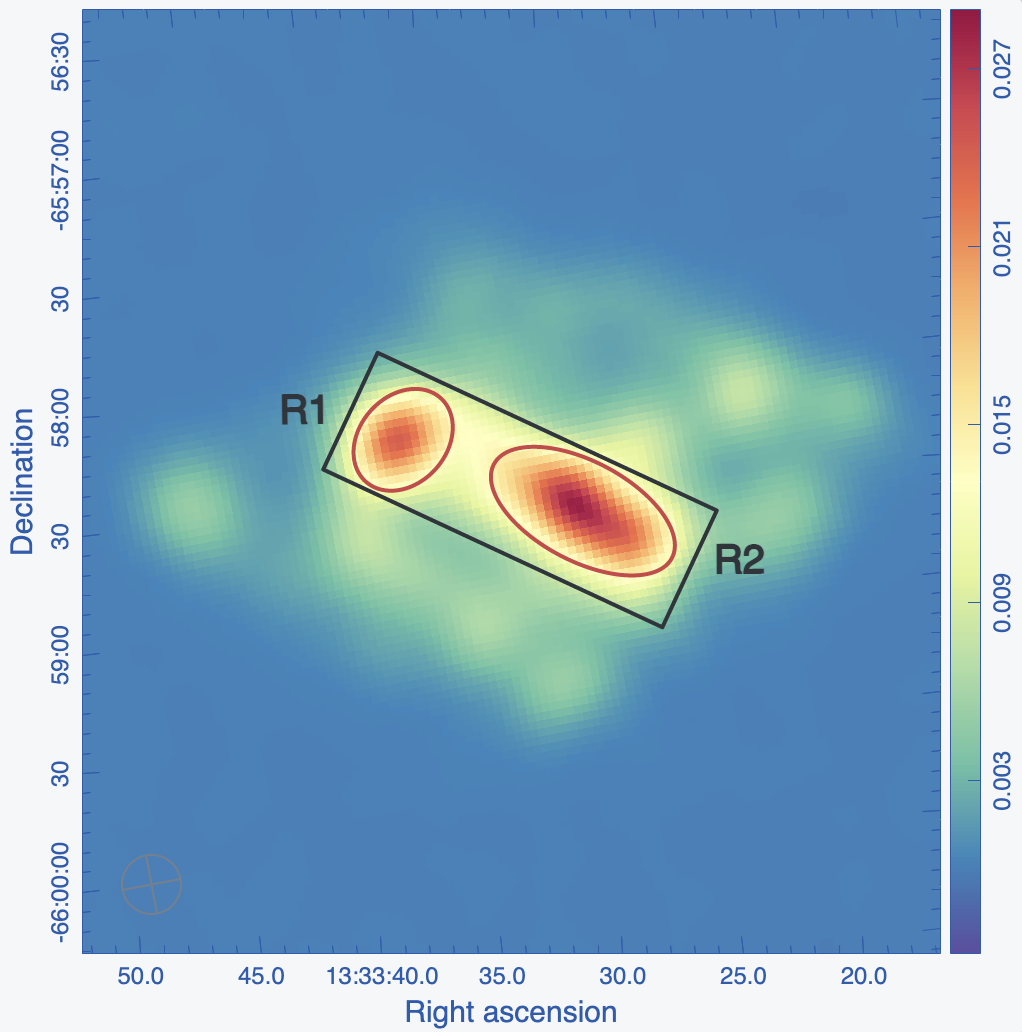}}
\caption{We outline in red the two inner envelopes in the central region of \emph{Infinity} (which we arbitrarily identify as regions R1 and R2), from which we measure the respective integrated flux densities. Additionally, we measure the apparent size of the inner region containing R1 and R2, as outlined by the central black rectangle.} 
\label{fig:inner_region}
\end{figure}

We use {\tt CARTA} and the polygon tool to define the outline of the two inner envelopes in \emph{Infinity's} central region, which we arbitrarily refer to as regions R1 and R2 in Figure~\ref{fig:inner_region} (measured in Jy/beam). The central coordinates for R1 are RA (J2000) 13:33:37.2 and DEC (J2000) --65:58:14, where we measured an integrated flux density of $S_{\rm 943\,MHz} =$ 0.0312\,Jy (SB53310) and $S_{\rm 943\,MHz} =$ 0.0315\,Jy (SB62225). The central coordinates for R2 are RA (J2000) 13:33:30.0 and DEC (J2000) --65:58:36, where we measure an integrated flux density of $S_{\rm 943\,MHz} =$ 0.0696\,Jy (SB53310) and $S_{\rm 943\,MHz} =$ 0.0698\,Jy (SB62225).

We measure the apparent size of the inner region containing only the R1 and R2 envelopes (indicated by the central black rectangle in Figure~\ref{fig:inner_region}), which is 1\farcm6\,$\times$\,0\farcm52, corresponding to a physical distance of 0.70\,pc\,$\times$\,0.23\,pc. Individually, the apparent size of the R1 envelope is 0\farcm48\,$\times$\,0\farcm37, corresponding to a physical distance of 0.21\,pc\,$\times$\,0.16\,pc. For the R2 envelope, we measure 0\farcm85\,$\times$\,0\farcm42, corresponding to a physical distance of 0.37\,pc\,$\times$\,0.18\,pc.

Using the mean of each of the R1 and R2 integrated flux density values, we also measure the optical depth and electron density values for these two inner envelopes, using the same methodology in equations \ref{eqnrayleigh} to \ref{EM} inclusive (this includes assuming a cylindrical geometry and hence measuring a path length equal to the smallest diameter, which for R1 is 0.16 pc and for R2 is 0.18 pc). This approach will determine whether \emph{Infinity} becomes optically thick in the line of sight to the central region, which would not only impact the overall integrated flux density measurement but possibly explain why the spectral index value is slightly higher than the optically thin case of $\alpha$ = --0.1. For R1 we measure $\tau$ = 0.0098 and $N_{e}$ = 405.82\,cm$^{-3}$, and for R2 we measure $\tau$ = 0.011 and $N_{e}$ = 402.66\,cm$^{-3}$. We can infer from both results of $\tau$ and $N_{e}$  that \emph{Infinity} is optically thin in the R1 and R2 regions. In terms of the slightly raised spectral index value, this warrants further investigation in future research, particularly at lower frequencies where the optical thickness may not be negligible in a specific direction.

\cite{Danehkar2018} mapped the excitation in the inner regions using \ac{HST} \ac{WFC3} imaging and describe the innermost structures as two low-ionisation envelopes (hence, spectroscopic studies predominantly detect emission lines such as [N\,{\sc ii}] and [O\,{\sc i}], \citealt{Mari2023}) surrounded by highly ionised environments, resulting from recent outbursts of the progenitor \ac{AGB} star. According to \cite{Miszalski2009}, these low--ionisation structures are a defining feature of post common--envelope PNe that surround a Wolf-Rayet star, such as \emph{Infinity}. 

\section{SUMMARY}\label{s:Summary}

Our motivation for this study has been to contribute radio continuum measurements for the well-known Galactic planetary nebula NGC\,5189 (which we refer to colloquially as \emph{Infinity}), using the \ac{ASKAP} radio telescope array. The \ac{ASKAP}-\ac{EMU} survey has observed \emph{Infinity} at 943\,MHz, from which we have measured the integrated flux density, spectral luminosity, surface brightness, physical diameters, optical depth and electron column density. We also measured the integrated flux densities of \emph{Infinity} in the \ac{SUMSS} and \ac{PMN} data, and combined those with other published radio measurements to determine the most up-to-date spectral index. We infer that \emph{Infinity} is a thermal (free--free) emitting nebula, is in agreement with previous studies, and consistent with the expectations of a PNe composed of ionised plasma. 

\paragraph{Acknowledgments}

We are grateful for the useful comments and suggestions made by Sanja Lazarevi\'c, Velibor Velovi\'c, Denis Leahy, and to our referee, which have improved the manuscript. This scientific work uses data obtained from Inyarrimanha Ilgari Bundara, the CSIRO Murchison Radio-astronomy Observatory. We acknowledge the Wajarri Yamaji People as the Traditional Owners and native title holders of the Observatory site. CSIRO’s ASKAP radio telescope is part of the Australia Telescope National Facility. Operation of ASKAP is funded by the Australian Government with support from the National Collaborative Research Infrastructure Strategy. ASKAP uses the resources of the Pawsey Supercomputing Research Centre. Establishment of ASKAP, Inyarrimanha Ilgari Bundara, the CSIRO Murchison Radio-astronomy Observatory and the Pawsey Supercomputing Research Centre are initiatives of the Australian Government, with support from the Government of Western Australia and the Science and Industry Endowment Fund. 

ADA is proudly supported by the Australian Government Research Training Program (RTP) Scholarship and CSIRO ATNF Space and Astronomy Student Program.

This research is also based on observations made with the NASA/ESA Hubble Space Telescope obtained from the Space Telescope Science Institute, which is operated by the Association of Universities for Research in Astronomy, Inc., under NASA contract NAS 5–26555. These observations are associated with program 12812.

\bibliography{MAIN_REFS}

\end{document}